\documentclass[usenatbib]{mnras}
\usepackage{newtxtext,newtxmath}
\usepackage[T1]{fontenc}
\usepackage{ae,aecompl}
\usepackage{graphicx, epsfig, graphics}	
\usepackage{amsmath}	
\usepackage{amssymb}	

\usepackage[left, switch, pagewise]{lineno}

\newcommand{\ergps}{erg s$^{-1}$}
\newcommand{\mpcc}{$m_{\rm p}$ cm$^{-3}$}
\newcommand{\kmps} {km s$^{-1}$}
\newcommand{\mpy}{M$_\odot$ yr$^{-1}$}

\newcommand{\tbf}{}

\title[NPS Asymmetry and Fermi Bubbles]{Possible connection between the asymmetry of North Polar Spur and Loop I with Fermi Bubbles}

\author[K. C. Sarkar]{
Kartick C. Sarkar\thanks{E-mail: \url{sarkar.kartick@mail.huji.ac.il}}
\\
Center for Astrophysics and Planetary Science, Racah Institute of Physics, The Hebrew University of Jerusalem, Israel
}
\date{Accepted XXX. Received YYY; in original form ZZZ}
\pubyear{2018}

\begin{document}
\label{firstpage}
\pagerange{\pageref{firstpage}--\pageref{lastpage}}
\maketitle

\begin{abstract}
The origin of North Polar Spur (NPS) and Loop-I has been debated over almost half a century and is still unresolved. Most of the confusion is caused by the absence of any prominent counterparts of these structures in the southern Galactic hemisphere (SGH). This has also led to doubts over the claimed connection between the NPS and Fermi Bubbles (FBs). I show in this paper, that such asymmetries of NPS and Loop-I in both X-rays and $\gamma$-rays can be easily produced if the circumgalactic medium (CGM) density in the southern hemisphere is only smaller by $\approx 20\%$ than the northern counterpart in case of a star formation driven wind scenario. The required mechanical luminosity, $\mathcal{L} \approx 4-5\times 10^{40} $ \ergps (reduces to $\approx 0.3$ \mpy including the non-thermal pressure) and the age of the FBs, $t_{\rm age} \approx 28$ Myr, are consistent with previous estimations in case of a star formation driven wind scenario. One of the main reasons for the asymmetry is the projection effects at the Solar location. Such a proposition is also consistent with the fact that the southern FB is $\approx 5^\circ$ bigger than the northern one. The results, therefore, indicate towards a possibility for a common origin of the NPS, Loop-I and FBs from the Galactic centre (GC). I also estimate the average sky brightness in X-ray towards the south Galactic pole and North Galactic pole in the \textsc{rosat}-R67 band and find that the error in average brightness is far too large to have any estimation of the deficiency in the southern hemisphere.
\end{abstract}

\begin{keywords}
Galaxy: -- centre, outflow, X-ray, gamma-ray
\end{keywords}


\section{Introduction}
North Polar Spur is the second largest structure in the sky that extends from Galactic longitude $l \approx 20^\circ$ to $-30^\circ$ and Galactic latitude $b \approx 10^\circ-70^\circ$ in the form of an arc with a thickness of $\sim 15^\circ$. This is encircled by another structure called the Loop-I feature that extends $\sim 10^\circ$ beyond the NPS in almost all directions. Both these structures are visible in X-rays and Loop-I in $\gamma$-rays  in northern Galactic hemisphere towards the centre of our Galaxy \citep{Berkhuijsen1971, Sofue1979, Snowden1997, Sofue2000}. 

Although there are faint indications of southern counterparts \citep[see Fig 13 of][]{Ackermann2014}, the absence of prominent signatures in the southern hemisphere has shadowed the truth behind the origin of these structures. Despite several claims by \cite{Sofue1977a, Sofue1984, Sofue1994, Sofue2000, Sofue2003, BlandHawthorn2003, Kataoka2013, Sarkar2015b, Sofue2016, Kataoka2018} that the NPS is `Galactic centre' phenomena, the origin of the NPS still remains debated even after half a century of the first discovery of these structures. The main reason is the absence of significant counterparts in southern hemisphere and a superposition of a nearby ($\sim 200$ pc) OB association, Sco-Cen along the line of sight. This has led a part of the scientific community to believe that the NPS/Loop-I are compressed shells collectively driven by several supernovae (SNe) from Sco-Cen OB association \citep{Berkhuijsen1971, Egger1995}. In this model, the apparent lack of the X-ray emission inner to the NPS is attributed to the absorption by a local HI shell around the local bubble. Although this HI shell would be sufficient to explain the absorption in the lower energy band ($0.1-0.4$ keV), it can not be the only reason for the lack of X-ray in $0.5-2.0$ keV band and indicates a true lack of X-ray emission in this region  \citep{Egger1995}.

Models, describing the NPS and Loop-I to be the emission arising from an interaction between two shells have been also proposed \citep{Wolleben2007}. The success of this model was to explain the polarised radio emission from the NPS/Loop-I and a `new loop' in the southern hemisphere. This new loop is also recognised by more recent observations by \cite{PlanckCollaborationXXV2016} as the `South Polar Spur'. The problem, however, stays in explaining the energetics of such shells. As noted by \cite{Shchekinov2018}, that the energy required to expand the HI shell generated at the interaction zone between these shells is equivalent to $\sim 60$ SNe (including the effect of cooling in the ISM). This is almost an order of magnitude larger than the expected number of SNe inside the Sco-Cen association over last $\approx 10$ Myr.

On the other hand, there are growing evidences that the NPS is not of a `local origin' and that its distance correlates well with the `Galactic centre origin' scenario. X-ray observations by \cite{Kataoka2013, Lallement2016} indicate that the NPS is highly absorbed by a hydrogen column density up to, $N_{\rm HI} \sim 4 \times 10^{21}$ cm$^{-2}$ towards the Galactic disc indicating a distance $\gg 200$ pc. Although, most of the volume within $150$ pc is occupied by the local bubble \citep{Egger1995}, it is, in principle, possible to achieve such a high column density within $\lesssim 200$ pc provided there is compressed wall of high density gas at $15-60$ pc region between the local bubble and the NPS \citep{Willingale2003a}. Although, observations by \cite{Lallement2014} indicate the presence of a high density shell towards the NPS, the required column density still falls short. This indicates the NPS to be beyond $\sim 4$ kpc \citep[see Fig 11 of][]{Lallement2016}

By analysing \textsc{O viii} Ly-$\alpha$ and Ly-$\beta$ and other Lyman series lines from \textit{Suzaku} and XMM-$Newton$ spectrum, \cite{Gu2016} also found that the lines are well explained if they are absorbed by a $0.17-0.20$ keV ionised medium with required hydrogen column density $N_{\rm H} \sim 5 \times 10^{19}$ cm$^{-2}$. This value is much more than what the local bubble could have provided ($\sim 5\times 10^{-3} \times 200$ cm$^{-3}$ pc $\approx 3 \times 10^{18}$ cm$^{-2}$). Moreover, the temperature of the local bubble ($\sim 10^6$ K; \citealt{Egger1995}) is also less than the required value. On the other hand, such temperature and column density for the absorption is easily achievable if the NPS is $\sim 8-10$ kpc into the CGM, assuming $T_{\rm CGM} \approx 0.2$ keV and density $\sim 10^{-3}$ \mpcc \citep{Henley2010a, Miller2015}. Another factor that goes against the NPS to have a local origin is its metallicity. Fitting of X-ray spectrum shows that the metallicity of the NPS is $\approx 0.3-0.7$ Z$_\odot$ \citep{Kataoka2013, Lallement2016} which is \tbf{closer} to the CGM value \citep[$\approx 0.5$ Z$_\odot$;][]{Miller2015, Faerman2017} than that of the local interstellar medium \citep[$\approx$ Z$_\odot$;][]{Maciel2010}.

The estimated density ($\approx 2\times 10^{-3}$ \mpcc), temperature ($\approx 0.25-0.35$ keV) and metallicity ($\approx 0.3-0.7$ Z$_\odot$) of NPS are suggestive of a structure in the Galactic CGM compressed by a Mach $\sim 1.5$ shock which could have been originated from the Galactic centre \citep{Kataoka2013}. This particular conclusion has far reaching implications towards understanding the origin of the Fermi Bubbles (FBs) as it directly constrains the energetics and thus the age of these bubbles and rules out many existing models. 

Since the discovery of the FBs \citep{Su2010} and further studies \citep{Ackermann2014, Keshet2016, Keshet2017}, there have been a numerous number of arguments regarding the origin of these bubbles. The arguments, can be classified into three main categories - (i) high luminosity ($\sim 10^{42-44}$ \ergps) wind driven by the central black hole \citep{Zubovas2011, Guo2012b, Zubovas2012, Yang2012, Yang2017} requiring the age of FBs to be $t_{\rm age} \sim$ few Myr, (ii) low luminosity ($\sim 2\times 10^{41}$ \ergps) wind driven by accretion disc around the central black hole, with $t_{\rm age} \approx 12$ Myr \citep{Mou2014, Mou2015} and (iii) star formation driven wind (star formation rate $\approx 0.1-0.3$ \mpy) with estimated age of $\approx 25-300$ Myr \citep{Crocker2015, Sarkar2015b}. There are also other constrains from \textsc{O viii} to \textsc{O vii} line ratio towards the FBs that are in favour of option ii and iii \citep{Miller2016b, Sarkar2017}. Such a variety of arguments crucially depend on whether one considers the NPS, Loop-I and FBs to be a `common origin' or not and would collapse to a small parameter space if one can answer the very origin of the NPS and Loop-I. 

Despite a number of arguments regarding the NPS/Loop-I to be a Galactic centre (GC) phenomena, their origin is still questioned and revolves around the fact that these structures are asymmetric across the Galactic disc. Interestingly, \cite{Kataoka2018} speculates that such an asymmetry could have been originated from an asymmetric density in the southern hemisphere. However, the fact that both the northern and southern FBs are almost of same size lead them to conclude that the NPS and Loop-I are probably a result of previous star-burst episode of the GC. In this paper, I show that a common origin for NPS, Loop-I and FBs is possible and that the asymmetric nature of the NPS and Loop-I can be obtained by having a local asymmetry in the CGM density without affecting the symmetry of FBs.

The base of the arguments, presented in this paper, crucially depend on the projection effects of the large scale structures. It has been shown by \cite[][hereafter SNS15]{Sarkar2015b} that the NPS/Loop-I are the outer shock (OS) of a star formation driven wind and has reached a distance of $\approx 8$ kpc starting $\approx 27$ Myr ago from the GC. Since we are at $\approx 8.5$ kpc away from the GC, the projection effects put this OS at $b \sim 70^\circ$ and $l \sim 60^\circ$. Now, if the CGM density in the southern hemisphere is slightly lower then the OS, in that hemisphere, has just run past the Solar system and, therefore, does not appear have a clear signature like a shock. The FBs, if considered to be the contact discontinuity (CD), then \tbf{do} not have to be very asymmetric. This would solve the tension between an asymmetric NPS/Loop-I and symmetric FBs as feared by \cite{Kataoka2018}.

The rest of the paper provides full details of the above arguments and presents numerical studies in a realistic Galactic environment generating X-ray, $\gamma$-ray and radio sky maps that can be compared with the actual observations from \textsc{rosat} and \textit{Fermi Gamma-ray Space Telescope}. 

\section{Numerical set up}
\label{sec:numerical-set-up}
This problem is studied by performing hydrodynamical simulations, without magnetic field and cosmic rays (CR). The simulations are performed using \textsc{pluto-v4.0} \citep{Mignone2007}. 
Since the shock structure crucially depends on the exact density distribution, we pay careful attention to the initial numerical set up. This set up is exactly the same one as presented in \cite[][hereafter SNS17]{Sarkar2017} (which was adapted from \cite{Sarkar2015a} to represent our Galaxy) except \tbf{for} a few modifications. 

%
\begin{figure*}
	\centering
	\includegraphics[width=\textwidth]{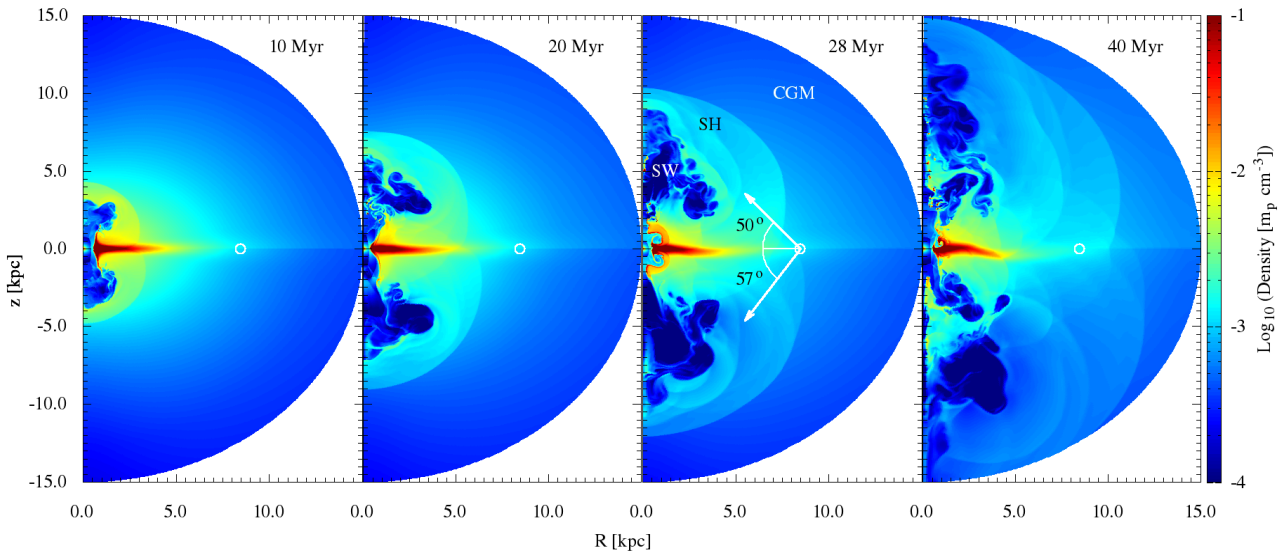}
	\caption{Evolution of the density contours for $f_h = 1/3$ . The Solar location has been shown by the white circle at $R, z = 8.5, 0.0$ kpc. The arrows in the third panel represent the apparent hight of the contact discontinuity and therefore the height of the FBs at the present moment. SW: shocked wind, SH: shocked halo}
	\label{fig:density-evo}
\end{figure*}

\subsection{Initial condition}
\label{subsec:initial-condition}
In SNS17, we considered that the CGM ($T_{\rm CGM} = 2\times 10^6$ K) is in hydrostatic equilibrium with the background gravity of dark matter, stellar disc and bulge. The parameters for the gravity and CGM temperature were fixed to best match the observed values. The resultant density distribution of the CGM was found to mimic the inferred density distribution from the \textsc{O viii} and \textsc{O vii} line emissions. I have, however, introduced some modifications to SNS17 set up to make it suitable for the present study.

A warm ($\approx 5\times 10^4$ K) and dense ($\sim 1$ \mpcc) gaseous disc in the initial density distribution has now been introduced. The disc gas is assumed to be rotating at $97.5\%$ level of the rotation curve. The rest of the support against gravity is provided by the thermal pressure. I also introduce a rotation to the hot CGM to comply with the observations of \cite{Miller2016a}. The speed of rotation, however, is assumed to be only a fraction ($f_h = 1/3$) of the Galactic disc rotation at that cylindrical radius ($R$). Although, \cite{Miller2016a} find that the CGM rotation speed is $\sim 180$ \kmps, their assumption of a spherical gaseous distribution makes this value uncertain. A proper estimation of the CGM rotation would require self consistent consideration of the flattening of the CGM arising due to rotation. Since that is not the main focus of this paper, I consider $f_h$ to take different values ($1/3, 1/2$ or $2/3$) to make up for this caveat. 
The exact pressure and hence the density distribution is then obtained by assuming that the disc gas and the CGM are both in steady state equilibrium with the background gravity  \citep[see][for details]{Sarkar2015a}.

I also switched off radiative cooling for $|z| \leq 1$ kpc to avoid artificial radiative cooling in the disc.
An active cooling in this region would \tbf{cause} the numerical disc to collapse into a thin layer of cold gas. In reality, turbulence generated by SN activity and infalling gas are responsible for maintaining  a fluffy disc \citep{Krumholz2017a}.  Since the current set up does not contain any of these physics, switching off the cooling in the disc is a way around this issue. As mentioned earlier, the NPS/Loop-I or FBs are structures in the CGM, therefore, this implementation is not expected to affect the results. Relaxing this assumption would lead to a large amount of injected energy to be lost via radiation within first $2-3$ Myr. This radiation loss would, however, reduce sharply as the shock breaks out of the ISM and starts propagating into the CGM. Therefore, the required energy would be $\sim 10\%$ ($\sim 2$ Myr$/28 $ Myr) higher than the injected value to achieve similar dynamics.

To achieve the purpose of this work, I assume that the CGM density in the southern hemisphere is $20\%$ less than the northern counterpart. Since this introduces a pressure imbalance across the galactic disc, I further set the temperature of the southern CGM $20\%$ higher than the northern one. This temperature asymmetry is not very realistic.
It makes the SGH appear brighter than they would be without such a temperature asymmetry. However, the actual brightness difference between the north and south (without the FBs) depends on the size of the asymmetric region and \tbf{is} discussed more in Section \ref{sec:hint-observation}.

The density asymmetry in the CGM can be caused either due to the motion of our Galaxy through the local group that caused an asymmetric ram pressure on the CGM or some previous star formation driven wind activity. Although the motion of our Galaxy towards the centre of the local group (somewhat similar direction towards the Andromeda galaxy) is almost parallel to the Galactic disc \cite{VanDerMarel2012a}, a local density asymmetry of size $\sim 10$ kpc can still be present in the CGM  (see Section \ref{sec:hint-observation}).

\subsection{Grid and energy injection}
\label{subsec:grid-energy}
The computational box is chosen in 2D spherical coordinates which, by definition, assumes axisymmetry. The box extends till $15$ kpc in radial direction and from $0$ to $\pi$ in the $\theta$-direction. A total of $1024\times 512$ grid points were set uniformly in radial and $\theta$-direction\tbf{s}.  The resolution of the box is, therefore, $\approx 15\times 6$ pc$^2$ at $r = 1$ kpc and $\approx 15\times 61$ pc$^2$ at $r = 10$ kpc. Both the boundaries in the $r$-direction are set to be outflowing, whereas, the $\theta$ boundaries are set to be axisymmetric (i.e. only $v_\theta$ and $v_\phi$ is reversed). 

Supernovae energy is added within central $100$ pc \footnote{This value is somewhat arbitrary and is a typical for star forming regions. This particular choice, however, does not have much influence on the size of the OS or the contact discontinuity. It slightly affects the shape of  FBs as can be seen in Figure A1 of \cite{Sarkar2017}.} in the form of thermal energy. A constant mechanical luminosity is provided assuming a constant star formation rate (SFR) based on a Kroupa/Chebrier IMF and \textit{starburst99} \citep{Leitherer1999} recipe. The mass and energy injection rates are, thereafter, given by
\begin{eqnarray}
\dot{M}_{\rm inj} &=& 0.1\,\, {\rm SFR} \nonumber \\
\mathcal{L} &=& 10^{41} \times \frac{\rm SFR}{\mbox{\mpy}} \,\,\, \mbox{\ergps}
\label{eq:mech-lumn-sfr}
\end{eqnarray} 
where, only $30\%$ of the SNe energy is assumed to survive the interstellar radiation loss in the initial SN expansion phase and become available for driving a large scale wind. 

In all the simulations presented here, I assume a mechanical luminosity $\mathcal{L} = 4.5 \times 10^{40}$ \ergps which was found to match the observed X-ray and $\gamma$-ray signatures of the NPS and the FBs in SNS15. If converted directly to SFR, this luminosity would imply SFR $\approx 0.45$ \mpy. We, however, should keep in mind that the non-thermal components like magnetic field and cosmic rays can contribute a large fraction of this energy and, therefore, the required star formation rate would decrease further from this value as noted in SNS15.

\section{Results and Discussion}
The evolution of density for $f_h = 1/3$ has been shown in Figure \ref{fig:density-evo}. As can be seen, the structure of the outflowing gas is similar to a wind driven shock as studied by \cite{Castor1975, Weaver1977}. In the inner part, it contains a free wind region which undergoes a reverse shock shortly. The wind material extends till the contact discontinuity (CD) beyond which the shocked CGM continues till the OS. Since the mass injected by the SNe driven wind is very small, the region inside the CD has low density gas (compared to the background CGM) which makes it suitable for hosting a X-ray cavity. Note that the $\gamma$-ray or radio emission, on the other hand, depend on the CR energy density which is dependent on the presence of shocks and turbulence. Given that there is a reverse shock (Mach $\sim 10$) and a turbulent medium inside, it is likely that this region hosts high energy cosmic ray electrons and, therefore, produce the observed FBs or the microwave haze. \tbf{These} arguments were used by \cite[][SNS15]{Mertsch2011} to assume that the FBs can be represented by the inner bubble extending all the way till the CD. Due to the lack of cosmic ray physics in the current simulations, I also follow the same arguments. While this argument is persuasive and likely true, a better understanding should, in any case, be built by performing numerical simulations including both CR physics and magnetic field.

\begin{figure}
	\centering
	\includegraphics[width=0.45\textwidth, angle=-90, clip=true, trim={0 3.5cm 0 4cm}]{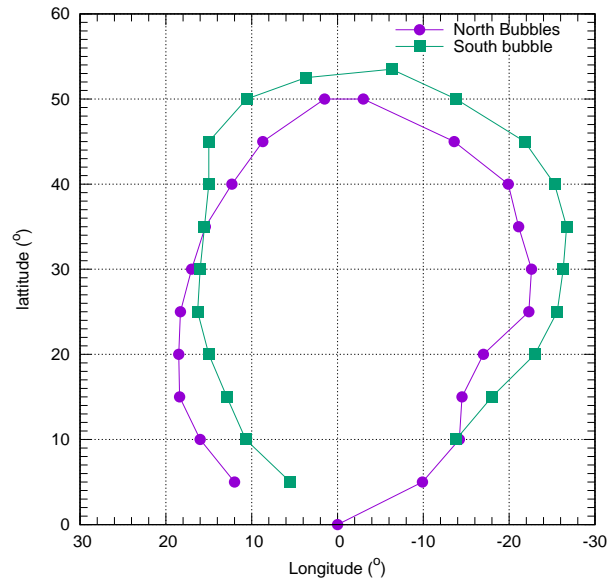}
	\caption{The outer edge of the FBs taken from \protect\cite{Su2010}. The southern bubble has been inverted in latitude to compare it with the northern bubbles. There is a clear signature that the southern bubble is $\approx 5^\circ$ bigger than the northern one. }
	\label{fig:bubbles}
\end{figure}

Based on the above arguments, the age of FBs is the time when the CD reaches $\approx 50^\circ$, i.e. $t_{\rm age} \approx 28$ Myr as can be seen in the third panel of Figure \ref{fig:density-evo}. Note that, here $t_{\rm age}$ is taken when the northern FB reaches $50^\circ$ (observed size of the northern FB). The southern bubble, however, appears to be $\approx 7^\circ$ bigger in latitude. It is indeed interesting to note that although both the observed FBs are considered to be of similar size, a careful look at these bubbles reveal that the southern bubble is $\approx 5^\circ$ bigger than the northern one. In Figure \ref{fig:bubbles}, I re-plotted the outer edge of the FBs (taken from \citealt{Su2010}) to establish this point. The figure shows a consistently larger southern bubble in all directions except in the bottom left part which may occur due to local density variation. 

\begin{figure}
	\centering
	\includegraphics[width=0.5\textwidth, clip=true, trim={6cm 2cm 5cm 4cm}]{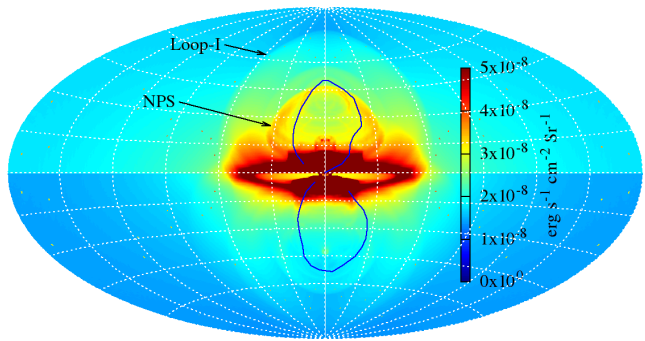}
	\caption{Soft X-ray ($0.5-2.0$ keV) sky map in Aitoff projection generated at $t_{\rm age} = 28$ Myr. NPS and Loop-I like features are clearly visible in the northern hemisphere compared to southern hemisphere. The surface brightness compares well with the observed values in \textsc{rosat} R6R7 band. The outer edge of the FBs have been over-plotted (blue solid lines) for visual comparison. A uniform background of $5 \times 10^{-9}$ \ergps cm$^{-2}$ Sr$^{-1}$ (comparable to $\approx 10^{-4}$ counts s$^{-1}$ arcmin$^{-2}$ in \textsc{rosat} R6R7 band) has been added to account for the observed diffuse X-ray background/foreground.}
	\label{fig:x-ray-map}
\end{figure}

\subsection{X-ray sky map}
\label{subsec:X-ray-map}
As mentioned in earlier discussion, the projection effects are very important while comparing simulations with observations of large structures in our Galaxy. I have made use of the module \textsc{pass}\footnote{Projection Analysis Software for Simulations (\textsc{pass}) described in SNS17. This code is freely available at \url{https://github.com/kcsarkar/}.} to produce proper projection effects at the Solar location. 

Figure \ref{fig:x-ray-map} shows \tbf{a} $0.5-2.0$ keV X-ray sky map generated at $t_{\rm age} = 28$ Myr from simulations \footnote{An extended box of $200$ kpc is also included to account for the emission beyond the computational box. The density asymmetry in the SGH, however, iss considered only till $50$ kpc.} with CGM rotation of $f_h = 1/3$ \footnote{See appendix for maps with CGM rotation of $f_h = 1/2$ and $2/3$.}. It shows the presence of features very similar to the NPS and Loop-I in the northern hemisphere along with the absence of these features in the southern part. A lower density in the southern hemisphere affects the surface brightness in two ways. Firstly, a $20\%$ lower density means a $\sim 40\%$ drop in X-ray brightness since the emissivity is $\propto n^2$. Secondly, due to a lower density the shock runs faster in the southern part and at $t = t_{\rm age}$, the OS just crossed us while the northern shock is still in front of us. Once we are inside shock, the projection effect makes it hard for us to detect any such shock in the southern hemisphere. 

The NPS, as seen in the current simulations, is not simply the shell that extends from the CD to the OS (in contrast to what was seen in SNS15). As can be noticed in Figure \ref{fig:density-evo},  there are few shocks present between the CD and the OS. Although, the presence of these shocks are not expected from simple analytical considerations, they arise due to the presence of an inhomogeneous and anisotropic medium and a low luminosity wind. For a typical wind scenario where the luminosity is very high, the wind is able to overcome the effect of disc pressure and thus follow a standard wind structure. However, for a low luminosity wind where the oblique ram pressure of the wind is just larger than the disc pressure, the free wind gets nudged at certain moments and thus produce\tbf{s} a variable luminosity wind. The shocks between CD and OS are generated due to such nudging. Note that this is also a channel by which the disc material gets entrained by the free wind and can produce high velocity warm clouds \cite{Sarkar2015a}.

The NPS is, therefore, the projection of one such shock close to the CD and does not have to be extended till the Loop-I (See third panel of Fig \ref{fig:density-evo}). While such a shock  follows the CD, it does not necessarily follow the outer edge of the $\gamma$-ray emission (as can be understood from Fig. \ref{fig:x-ray-map} and \ref{fig:g-ray-map}). We speculate that such secondary shocks may also be the origin of the \textit{inner arc} and \textit{outer arc}. It is also possible that the shock edge detected in \cite{Keshet2017} could be one of such secondary shocks.

\subsection{$\gamma$-ray sky map}
\label{subsec:gamma-map}
To generate the $\gamma$-ray map, I follow SNS15 and  assume that the main source of the $\gamma$-ray emission is via inverse Compton of cosmic microwave background by high energy CR electrons (CRe) \footnote{It was shown in SNS15 that a hadronic process is ineffective in producing enough surface brightness for the FBs.} and that the total CR energy density if assumed to be $15\%$ (of which only $0.0075\%$ is in the CRe) of the local thermal energy density at any grid location. The CRe spectrum inside the FBs (in this case, the CD) is assumed to be $dN/dE \propto E^{-2.2}$ \citep{Su2010, Ackermann2014}, which is also the electron spectrum required for explaining the microwave haze \citep{PlanckCollaboration2013}. It is, therefore, generally believed that both the radio and $\gamma$-ray emission originated from the same population of CRe. Since such high energy CRe is expected to cool down via inverse Compton and synchrotron emission, a break at Lorentz factor $\Gamma = 2\times 10^6$ is also assumed. After this break the CRe spectrum follows $dN/dE \propto E^{-3.2}$. 

Outside the CD and inside the OS, a softer CRe spectrum is assumed, $dN/dE \propto E^{-2.4}$ and a break at $\Gamma = 2\times 10^6$ is considered. This spectrum is consistent with the estimated value for the Loop-I \citep{Su2010} although the break location and the cut-off frequency is somewhat uncertain . A softer spectrum is indeed  expected at the OS as it is much weaker (Mach $\sim 1.5$) than the reverse shock inside the FBs. 
Note that the above prescribed assumptions to get $\gamma$-ray emission are very simplistic. A better approach requires a self-consistent implementation of the evolution of CR spectra in real time. Our focus in this section is, however, only to show the size and shape of the FBs and Loop-I.

\begin{figure}
	\centering
	\includegraphics[width=0.5\textwidth,  clip=true, trim={6cm 2cm 5cm 4cm}]{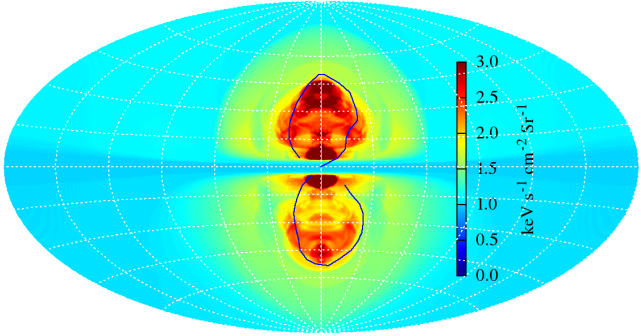}
	\caption{$\gamma$-ray sky map at $5$ GeV generated at $t_{\rm age} = 28$ Myr. The shapes and sizes of north and south FBs are consistent with the observed ones (blue solid lines). A clear shock structure is seen in the northern hemisphere but not in the southern part. A constant background/foreground of $1$ keV s$^{-1}$ cm$^{-2}$ Sr$^{-1}$ \citep{Su2010} is added in order to account for the observed diffuse emission. Also, regions with $|z|\leq 700$ pc is not included in the map to avoid any disc emission.}
	\label{fig:g-ray-map}
\end{figure}

Figure \ref{fig:g-ray-map} shows the  $\gamma$-ray sky map at $5$ GeV, generated at $t = 28$ Myr for CGM rotation of $f_h = 1/3$. It shows a good match for the size and shape of the FBs although the surface brightness inside the FBs is not as uniform as the observed ones. This can be attributed to the simple assumption of CR energy density to be a constant fraction of the thermal energy density. In reality, the CR behaves as a relativistic fluid (adiabatic index $= 4/3$) and, therefore, does not exactly follow the Newtonian plasma (adiabatic index $= 5/3$). Besides, the CR diffusion and the effect of the magnetic field in CR propagation is also not taken into account in the current numerical simulations.  Diffusion can make the CR energy density more uniform than the thermal pressure, while the inclusion of magnetic field can make the outer edge of the simulated FBs much smoother than as seen in observations.

Similar to the X-rays, a larger structure beyond the FBs is also noticed. This may correspond to the Loop-I structure seen in the northern sky. The surface brightness and contrast with the background seem to match quite consistently. However, an excess emission beyond the southern FB can be noticed, although no shock structure is clearly identifiable. This excess emission is in contrast with the observations. However, we should remember that, in the souther hemisphere, we are inside the shock and, therefore, the observable CRe spectrum is not the same as the northern Loop-I, it can be steeper due to lack of further re-acceleration of CRe behind the OS. This would mean that there is less amount of excess surface brightness in the southern hemisphere. For an example, the $\gamma$-ray emissivity for a $\propto E^{-2.45}$ CRe spectrum can be only $60\%$ of the emissivity for a $\propto E^{-2.4}$ spectrum, considering everything else to be the same. Therefore, it is possible that the excess brightness in the Southern part is not distinguishable from the background. At this point, it should be noted that although there is no clear signature of a southern counterpart of Loop-I, two rising $\gamma$-ray horns are clearly visible in the observations by \cite{Ackermann2014} (see their figure 13).

\subsection{Note on the East-West asymmetry}
\label{subsec:east-west-asym}
Along with a the North-South asymmetry, we also see an East-West asymmetry in NSP/Loop-I and also the Fermi Bubbles. While the NPS and the Loop-I are clearly seen to be bent towards the West, the bend in FBs are marginal but can be noticed in Fig \ref{fig:bubbles} (also see \citealt{Keshet2017}). In a shock propagation scenario, such \tbf{a deformation} is indication of an enhanced density towards the East. This also coincides with the fact that our `collision course' towards M31 is towards the East ($l \approx 121^\circ, b\approx 21^\circ$). The increased ram pressure from the intra-group medium can cause such density enhancement towards the East and, therefore, induce the observed East-West asymmetry. The fact that the NPS, Loop-I and the FBs are simultaneously bent towards the West (both in north and south for FBs) \tbf{is} another indication that these structures could be related to each other.

\begin{figure}
\centering
	\includegraphics[width=0.5\textwidth,  clip=true, trim={30cm 15cm 5cm 20cm}]{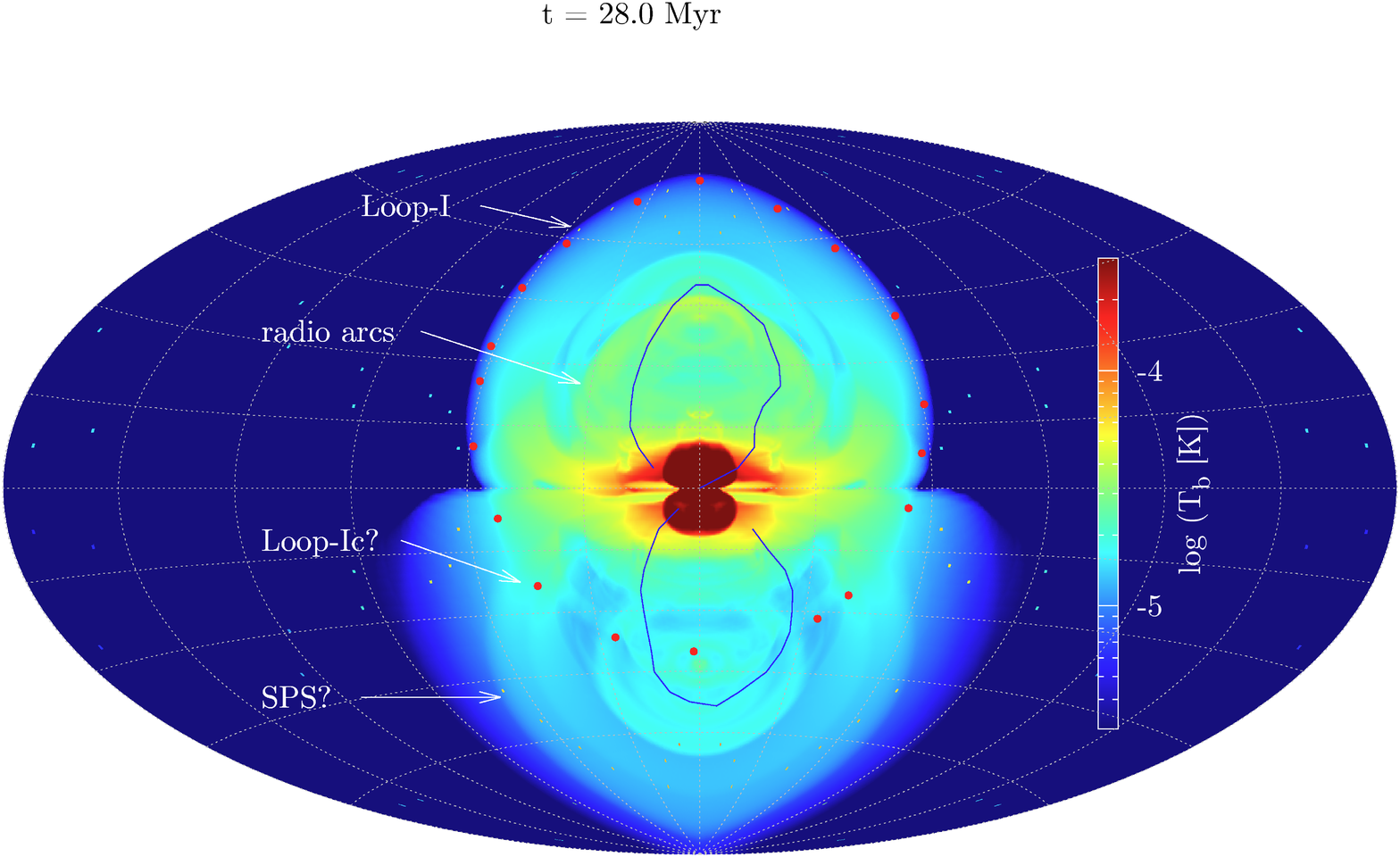}
	\caption{Brightness temperature map at 23 GHz. A possible location of the apparent southern counterpart of the Loop-I, called `Loop-Ic' here, is represented by the red dots to guide the eye.  A possible location of the SPS is also shown in the southern hemisphere. The appearance of the Loop-Ic has been discussed in Section \ref{subsec:radio-map} and the corresponding density map in shown in Figure \ref{fig:density-evo-cases}.}
	\label{fig:radio-map}
\end{figure}

\begin{figure*}
	\centering 
	\includegraphics[width=\textwidth]{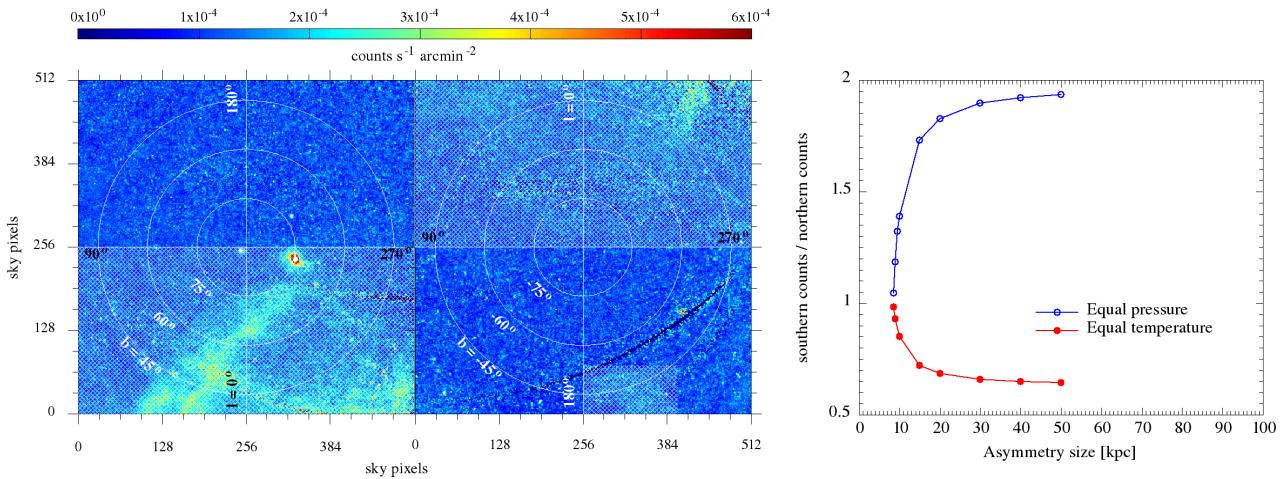}
	\caption{Estimating the asymmetry in X-ray sky brightness. Left panel: \textsc{rosat} R67 band sky brightness observed towards the north Galactic pole presented in Aitoff-Hammer equal area projection. Middle panel: R67 band sky brightness towards the south Galactic pole. The white hashed region represents the masks applied to avoid the area contaminated by the big scale structures, like the NPS  \citep{Snowden1997}.  Right panel: Expected asymmetry in the diffuse sky brightness towards the Galactic poles in \textsc{rosat}-R67 band. The blue curve shows a case, where the pressure was kept constant inside the cavity and the red curve shows the case where the temperature inside the cavity was kept constant. Data values above $10^{-3}$ counts s$^{-1}$ arcmin$^{-2}$ are removed from the plot to avoid very bright foreground}
	\label{fig:rosat-comp}
\end{figure*}

\subsection{Radio sky map}
\label{subsec:radio-map}
It has been argued several times that the Loop-I is a closed by ($\sim 200$ pc) feature and that there is a structure in the southern hemisphere that makes the Loop-I to be a complete loop in radio \citep{PlanckCollaborationXXV2016, Dickinson2018, Liu2018}. However, it should be noted that there is no unique way to draw such a loop as it is mostly driven by eye. One important feature, in this context, is the `South Polar Spur' (SPS) \cite[as named by][]{PlanckCollaborationXXV2016} or the `new loop' \cite[as named by][]{Wolleben2007}. Although it was initially thought to be part of a bigger loop, called Loop II, it was later ruled out as the curvature of the SPS is in the opposite direction of the Loop II. Interestingly, the curvature of the SPS is also bent towards the west much like the Loop-I, moreover, the magnetic field (MF) is also almost parallel to the structure \citep[see Fig 20 of][]{PlanckCollaborationXXV2016}.

Since the simulations presented in this paper do not have MFs, it is not possible to show the orientation of the MF. It can only be speculated that such fields would be compressed in the outer shock and also would be parallel to the shocked shell (much like a SN shell in the ISM). Approximate radio intensity maps can, however, be obtained with the help of some assumptions regarding the CR and the MFs energy density. Following SNS15, I assume that the CR and MF energy densities are only a fraction ($\epsilon_{\rm cr}$ and $\epsilon_B$, respectively) of the thermal energy ($u_{\rm th}$) at that location. Assuming that the synchrotron  emission is coming from an electron distribution of $n(E)\, dE = \kappa\, E^{-2.4}\, dE$ (as observed in Loop-I), the synchrotron emissivity per unit solid angle can be written as \citep{Longair1981}
\begin{equation}
J_\nu = 2.2\times 10^{-19}\, \epsilon_{\rm cr}\,\epsilon_B^{0.85}\, u_{\rm th}^{1.85}\, \nu_{\rm GHz}^{-0.7}\,\,\,\,\, \mbox{\ergps cm}^{-3} \mbox{Hz}^{-1}\mbox{Sr}^{-1} 
\end{equation}
Fig. \ref{fig:radio-map} shows the obtained $23$ GHz brightness temperature (in excess to the CMB) assuming $\epsilon_{\rm cr} = 0.15$ and $\epsilon_{B} = 0.4$ (as obtained in SNS15). The map clearly shows the Loop-I structure as well as many other radio structures that have similar arcs as the Loop-I and seem to originate from the centre, but lies at different longitudes. Similar but fainter signatures of such arcs are also seen in the southern hemisphere. Interestingly, it is possible to identify arcs in the southern hemisphere that are bent towards the west and are similar to the SPS in nature. It is also possible to draw an arc which could correspond to the southern part of the Loop-I but would be limited in extension.  Such a loop is shown as the Loop-Ic in Fig. \ref{fig:radio-map}. This arc is a part of one of the secondary shocks present in the simulation as mentioned in Section \ref{subsec:X-ray-map}. This arc has also been marked in Fig. \ref{fig:density-evo-cases} for convenience. It, therefore, appears that such a secondary loop can easily be mistaken as the southern counterpart of the Loop-I which would then lead to a shorter distance estimate for the structure. Note that the \textit{WMAP} Haze is not visible in the radio map as the assumed electron spectral index ($ x = 2.4$) is steeper than the spectral index in the Haze ($x = 2.2$).

\section{hints from the observations}
\label{sec:hint-observation}
\subsection{The data}
\label{subsec:the-data}
Although one can easily notice the asymmetry of sky brightness across the northern and southern hemispheres in the Fermi maps \citep[Fig 13 of ][]{Ackermann2014}, \textsc{rosat} maps \citep{Snowden1997} as well as in the emission measure along the FBs \citep[][]{Kataoka2015}, it is hard to compare the theoretical maps with the observations as the size of the NPS is different from observations. Moreover, due to the axisymmetric nature of the simulation, the mirrored NPS is also seen at $l\approx 330^\circ$. Also, there are few arcs, like the one from $b \approx 0 - 45^\circ$ at $l \approx 330^\circ$, that are present in the observed map of SGH but not in the theoretical maps. Therefore, it is hard to find out if there is truly any density asymmetry in the CGM by looking at the sky towards the Galactic centre ($270^\circ \lesssim l \lesssim 90^\circ$).

To find out if the initial density distribution was asymmetric, one needs to look at the region where there is no contamination from these structures, i.e. towards $270^\circ \gtrsim l \gtrsim 90^\circ$ . In an attempt to estimate the quantitative value of the proposed asymmetry, I consider the \textsc{rosat}-R67 data towards the north and south Galactic poles \citep[Fig 4a,b of][]{Snowden1997} which is corrected for point sources, exposure time and normalised to an effective `on-axis' response of the XRT/PSPC. Additionally, the expected absorption by neutral Hydrogen is minimal in this region of the sky and in this band ($0.73-2.04$ keV).

\subsection{Masks applied}
\label{subsec:masks}
I mask out regions within $270^\circ \leq l \leq 90^\circ$ to avoid any emission from extended structures that could have \tbf{been} generated from the forward shock as seen in Fig \ref{fig:x-ray-map}. The considered region by default excludes regions of $|b| \lesssim 30^\circ$ where the disc emission could be important. We also mask out a small region within $ 180^\circ \lesssim l \lesssim 210^\circ $ and $-45^\circ \lesssim b \lesssim -30^\circ$ where the emission seems to be related to a radio bright ($\sim 50 - 120 \mu$m in IRIS maps) arm extended from the disc. Moreover, its relatively sharp boundary makes it unsuitable for  studying the background CGM emission. The sky-maps and the masks applied for this estimation \tbf{are} shown in the left and middle panel of Fig \ref{fig:rosat-comp} by the white hashed region. Each of these panels shows a $102.4^\circ \times 102.4^\circ$ patch of the sky with a pixel size of $12'\times 12'$. The maps are shown in an Aitoff-Hammer equal area projection so that every pixel has the same area irrespective of its position in the sky.  I also remove some pixels ($3189$ in the north and $12511$ in the south) where i) the brightness is more than $10^{-3}$ counts s$^{-1}$ arcmin$^{-2}$ (to avoid any bright foreground emission) or ii) the data is missing or iii) the hardness ratio (R67/R45) was more than $3.0$ (to avoid any non-thermal emission). 

\subsection{results}
\label{subsec:obs-results}
After applying the above filters, the average brightness of the northern and southern patch is estimated to be $(11.7 \pm 6.7) \times 10^{-5}$ count s$^{-1}$ arcmin$^{-2}$ and $(11.2 \pm 7.3) \times 10^{-5}$ count s$^{-1}$ arcmin$^{-2}$, respectively. Although the ratio between the mean values indicate a $\sim 4\%$ deficiency in the southern hemisphere compared to the north, the error in estimating the ratio is $\sim 100\%$ (obtained from the error maps provided by \cite{Snowden1997}). Clearly, this result is unsuitable for putting any kind of constrain on the size of the asymmetric region. To have a more reliable data, one needs to model and subtract the effect of Solar flares and extragalactic sources from the data  in addition to having better sky maps from other X-ray missions. This is out of the scope of this paper.


In case, a better modelling of the non-CGM emission in the \textsc{rosat} data is able to minimise the error, the results can be compared to the right panel of Fig. \ref{fig:rosat-comp}. Here, I consider the initial set-up (with $f_h = 1/3$) as described in Section \ref{subsec:initial-condition} with different sizes of the asymmetric region ($r_{\rm asym}$). The temperature inside the cavity is assumed to be such that a) there is no pressure asymmetry and b) there is no temperature asymmetry across the north and south. The ratios between the average northern and southern sky brightness in R67 band ($0.73-2.04$ keV) in the region of  $90^\circ \leq l \leq 270^\circ$ and $|b| \geq 30^\circ$ are shown by the blue (equal pressure) and red (equal temperature) lines in this figure. The difference between these two cases arises due to the difference in emissivities at the assumed temperatures inside $r_{\rm asym}$. For equal pressure case, since the temperature inside the asymmetric region is assumed to be $20\%$ higher than the northern part, the emissivity inside $r_{\rm asym}$ is higher compared to the case where the cavity temperature is assumed to be the same as the background. We also notice that the ratio becomes $1$ for $r_{\rm asym} \simeq 9$ kpc. This is due to the fact that the effect of  such \tbf{a} smaller cavity would be visible only within $270^\circ \leq l \leq 90^\circ$ and is by default not considered in the estimation. 

In any case, with the current analysis of the \textsc{rosat} data,  it is hard to put any constrain on the size of the asymmetric region. A better constrain on the size should include better data and also an estimation of the temperature inside the proposed cavity. A similar exercise can also be done for the Fermi data to look for such an asymmetry. However, it involves very accurate modelling of the Galactic foreground and point sources in the $\gamma$-rays and is also out of the scope of this paper.

\section{Conclusion}
In this paper, I demonstrated the feasibility of an idea that the NPS, Loop-I and FBs can have a common origin despite the asymmetry of NPS/Loop-I across \tbf{the} Galactic disc and apparent symmetry between the FBs. I show that a density asymmetry, as small as $20 \%$, in the southern hemisphere can produce the sizes, shapes and surface brightness in X-ray and $\gamma$-ray along with the asymmetric signatures of the NPS and Loop-I strikingly similar to the observed ones. This asymmetry requires the southern FB \tbf{to be} only $\approx 7^\circ$ bigger than the northern one, which is consistent with the observations of the southern bubble to be $\approx 5^\circ$ bigger than the northern one. Note that this particular value of $20\%$ is only a choice to prove the feasibility of the idea. At this point, it is not very clear how such asymmetry could have originated. Best guesses are either from the motion of our Galaxy in the local group which caused an asymmetric ram pressure to the CGM or a previous activity of asymmetric SNe-driven wind.  Since the motion of the Milky Way towards the M31 is almost in a straight line (towards, $l \approx 121^\circ,\,b\approx 21^\circ$), it is more likely to produce an east-west asymmetry (as discussed in Section \ref{subsec:east-west-asym}) than a north-south asymmetry.  In such a case, the north-south density asymmetry is more likely to be generated from a previous episode of star formation that released more energy in the south than the north. We should however, keep this caveat of the proposed model in mind.

It is interesting to note that the same projection effects would also appear if the energy injection happens slightly below the Galactic mid-plane. In this case, the OS in \tbf {the} southern hemisphere would be given a head start compared to the northern part. Even then, the southern shell would still be visible in X-ray as the density of the shell is \tbf{the} same as the northern part. Besides, the observations already put the star forming region at the Galactic centre roughly at the mid-plane. Also, note that such a projection model works only in case of a SNe driven wind and not in AGN driven winds. As can be seen in Fig 4 of SNS17 that the AGN driven bubbles are more vertical and, therefore, are not expected to respond to such a projection effect at $t = t_{\rm age}$ as presented in this paper.

One concern is that the simulations are performed only at one mechanical luminosity $4.5\times 10^{40}$ \ergps that corresponds to SFR $\approx 0.45$ \mpy. This conversion, however, may change depending on the effect of non-thermal pressures, like the cosmic ray and the magnetic pressure. As seen in SNS15, the total non-thermal contribution is almost $50\%$ of the thermal contribution. This makes the required star formation rate $\approx 0.3$ \mpy for producing the above mechanical luminosity. This value is almost factor of $2-3$ higher compared to observed value of $\approx 0.1$ \mpy \citep{Yusef-Zadeh2009, Immer2012, Koepferl2015}. Moreover, the conversion from mechanical luminosity to the SFR also depends on the assumed thermalisation efficiency of the SNe. In Eq. \ref{eq:mech-lumn-sfr}, I assume this efficiency to be $0.3$ \citep{Gupta2016}. However, the calculation of such efficiencies are obtained when the bubble is expanding in an uniform medium and can be, in principle, higher than $0.3$ in case the bubble breaks out of the disk and expands in a \tbf{low density} medium where the radiation loss is negligible. Therefore, a higher thermalisation efficiency would mean a lower SFR to maintain the same mechanical luminosity.
A further limitation is the absence of proper CR physics and magnetic field in the simulations. This forces one to assume some prescriptions while calculating the non-thermal emission and also affects the conversion between the mechanical luminosity to SFR.

Despite these limitations, the potential of the arguments presented in this paper indicates towards a common origin of the NPS, Loop-I and the FBs. The current all sky maps that are available from \textsc{rosat} are unsuitable (due to large errors in the counts) for verifying the proposed scenario. Hopefully, future X-ray missions like, \textit{e-ROSITA} will be able to verify or nullify this scenario.

\section*{ACKNOWLEDGEMENTS}
It is a pleasure to thank Orly Gnat, Reetanjali Moharana, Biman Nath, Prateek Sharma, Yuri Shchekinov and Amiel Sternberg for helpful discussions and suggestions. I also thank the anonymous referee whose critical feedback improved the content of this article. This work was supported by the Israeli Centers of Excellence (I-CORE) program (center no. 1829/12), the Israeli Science Foundation (ISF grant no. 857/14) and DFG/DIP grant STE 1869/2-1 GE 625/17-1. I thank the Center for Computational Astrophysics (CCA) at the Flatiron Institute Simons Foundation, where some of the computations were carried out.




\appendix
\section{Effects of CGM rotation}
\begin{figure*}
	\centering 
	\includegraphics[width=0.95\textwidth]{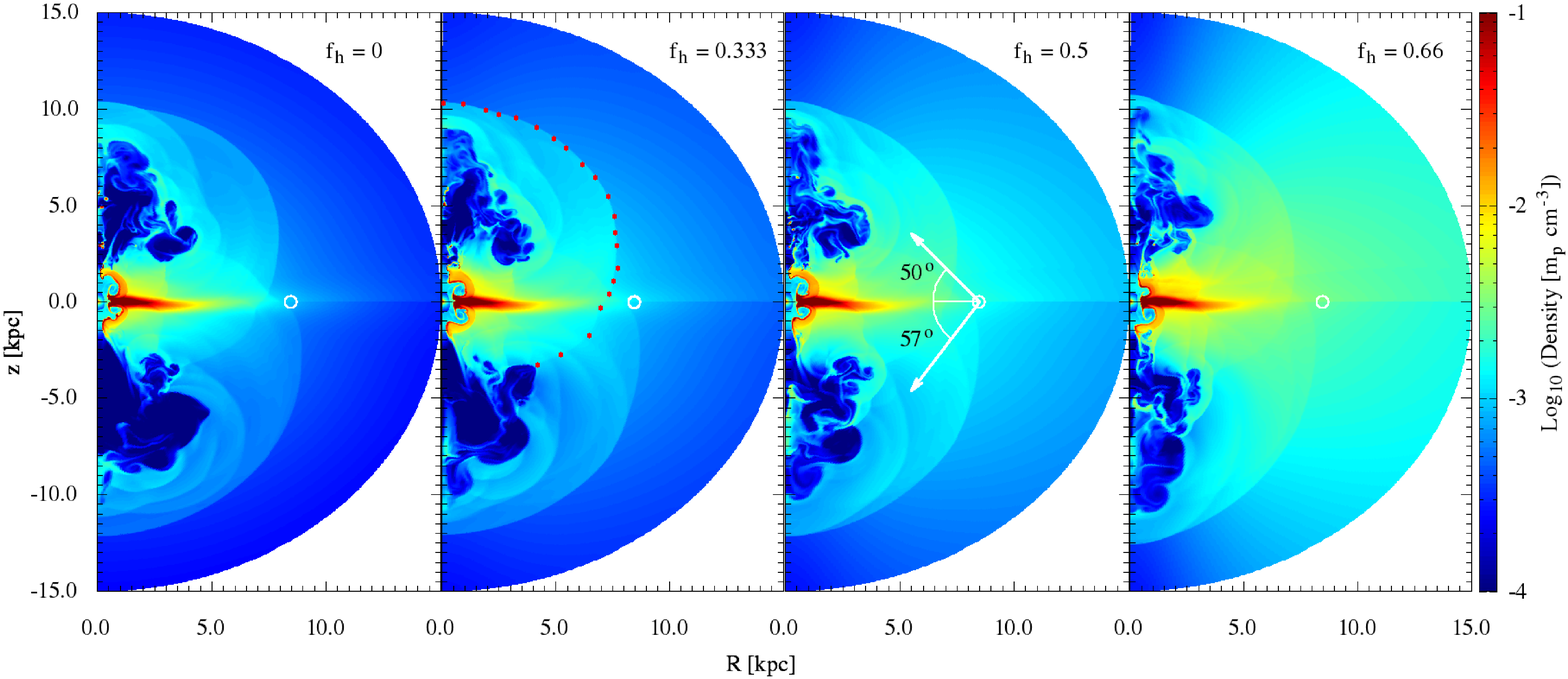}
	\caption{Density structure for simulations with different CGM rotation ($f_h$). The contours are plotted at $t_{\rm age} = 28$ Myr. With higher CGM rotation (higher $f_h$), the density becomes more flattened. The red dots in the second panel represent an apparent asymmetric (north-south) shock structure that could appear as the Loop-Ic in the radio map as discussed in Section \ref{subsec:radio-map}.}
	\label{fig:density-evo-cases}
\end{figure*}

\begin{figure}
	\centering
	\includegraphics[width=0.41\textwidth, clip=true, trim={6cm 2cm 6cm 3cm}]{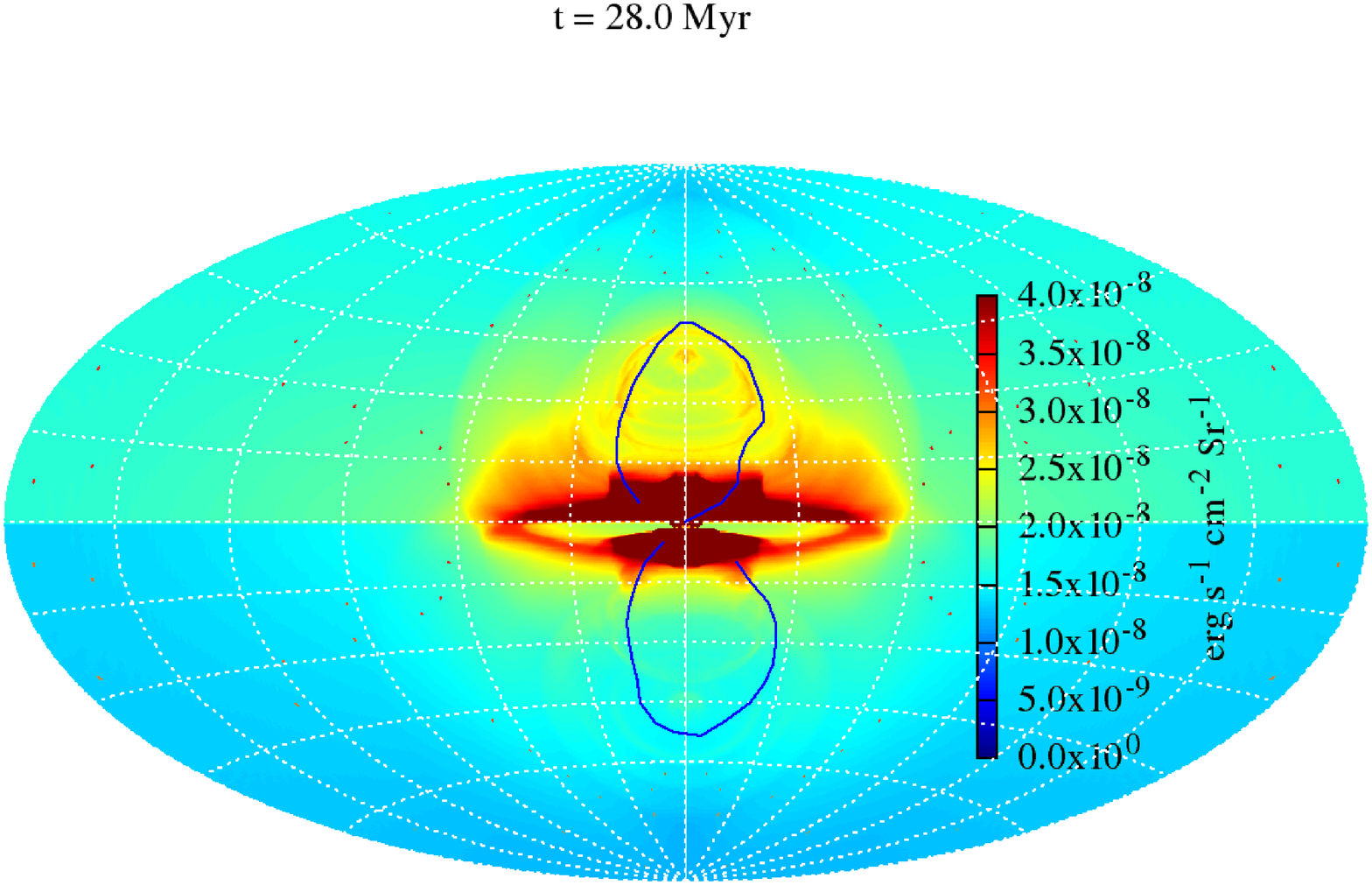}
	\includegraphics[width=0.41\textwidth, clip=true, trim={6cm 2cm 6cm 3cm}]{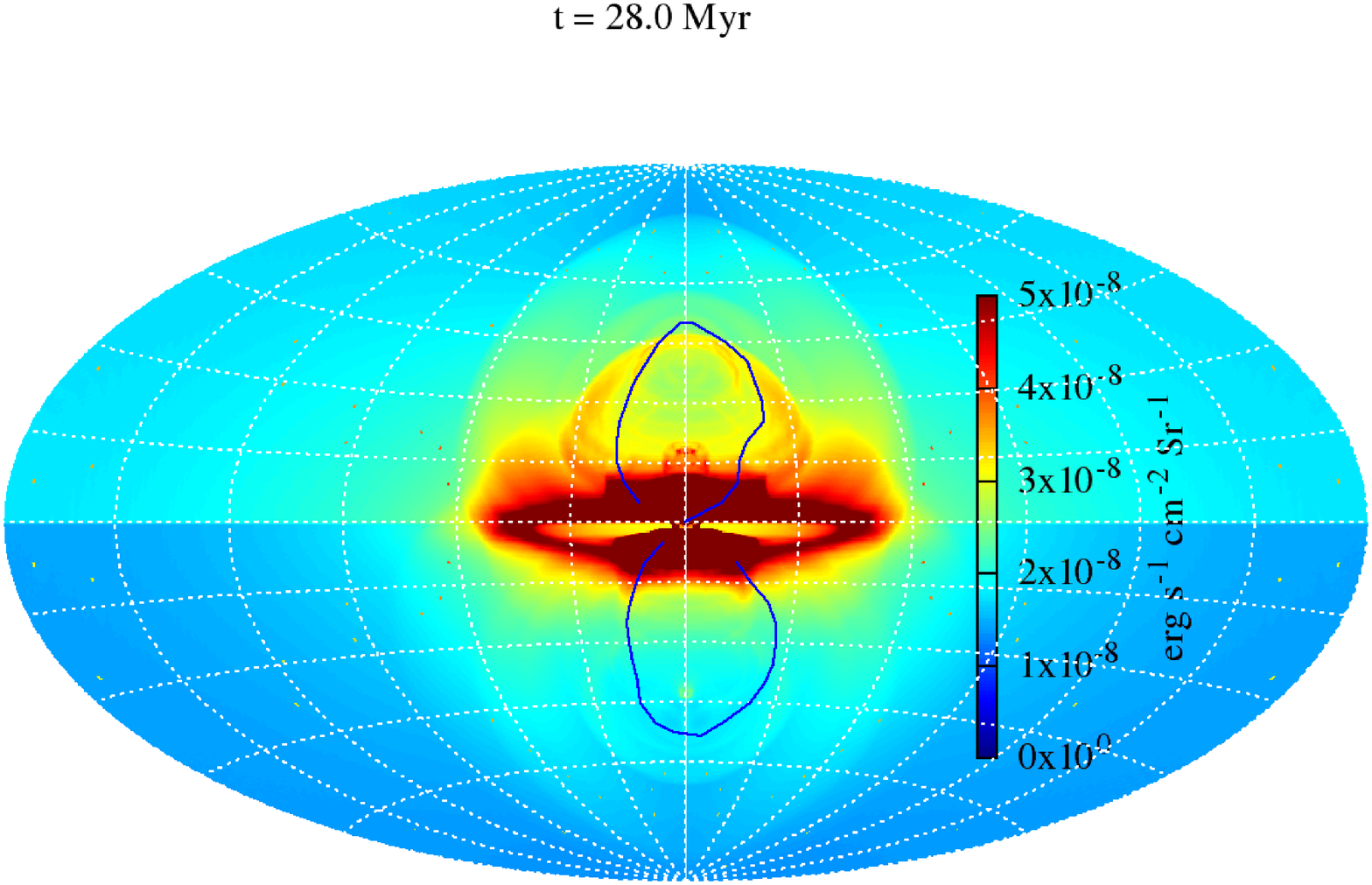}
	\includegraphics[width=0.41\textwidth, clip=true, trim={6cm 2cm 6cm 3cm}]{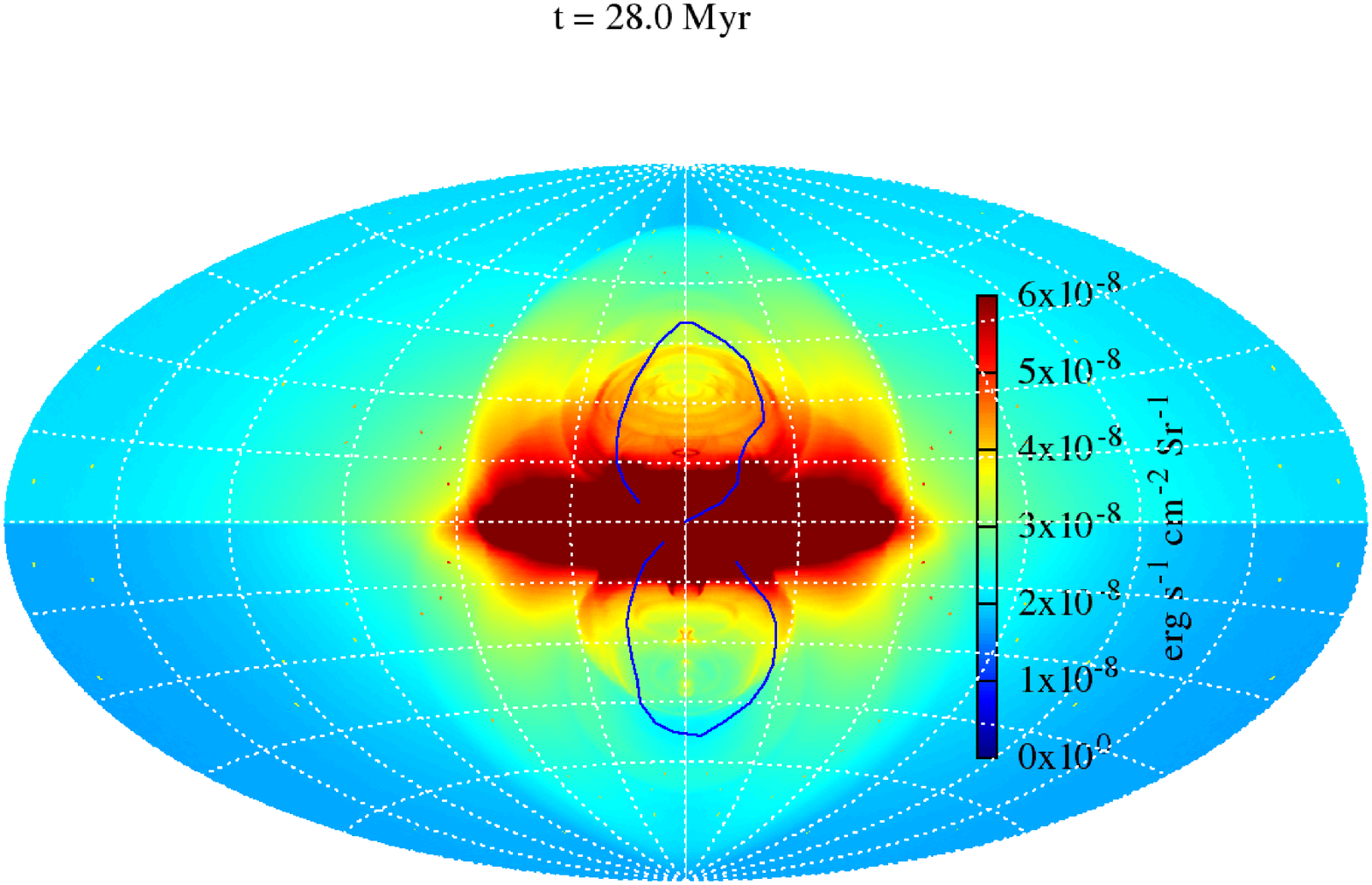}
	\includegraphics[width=0.41\textwidth, clip=true, trim={6cm 2cm 6cm 3cm}]{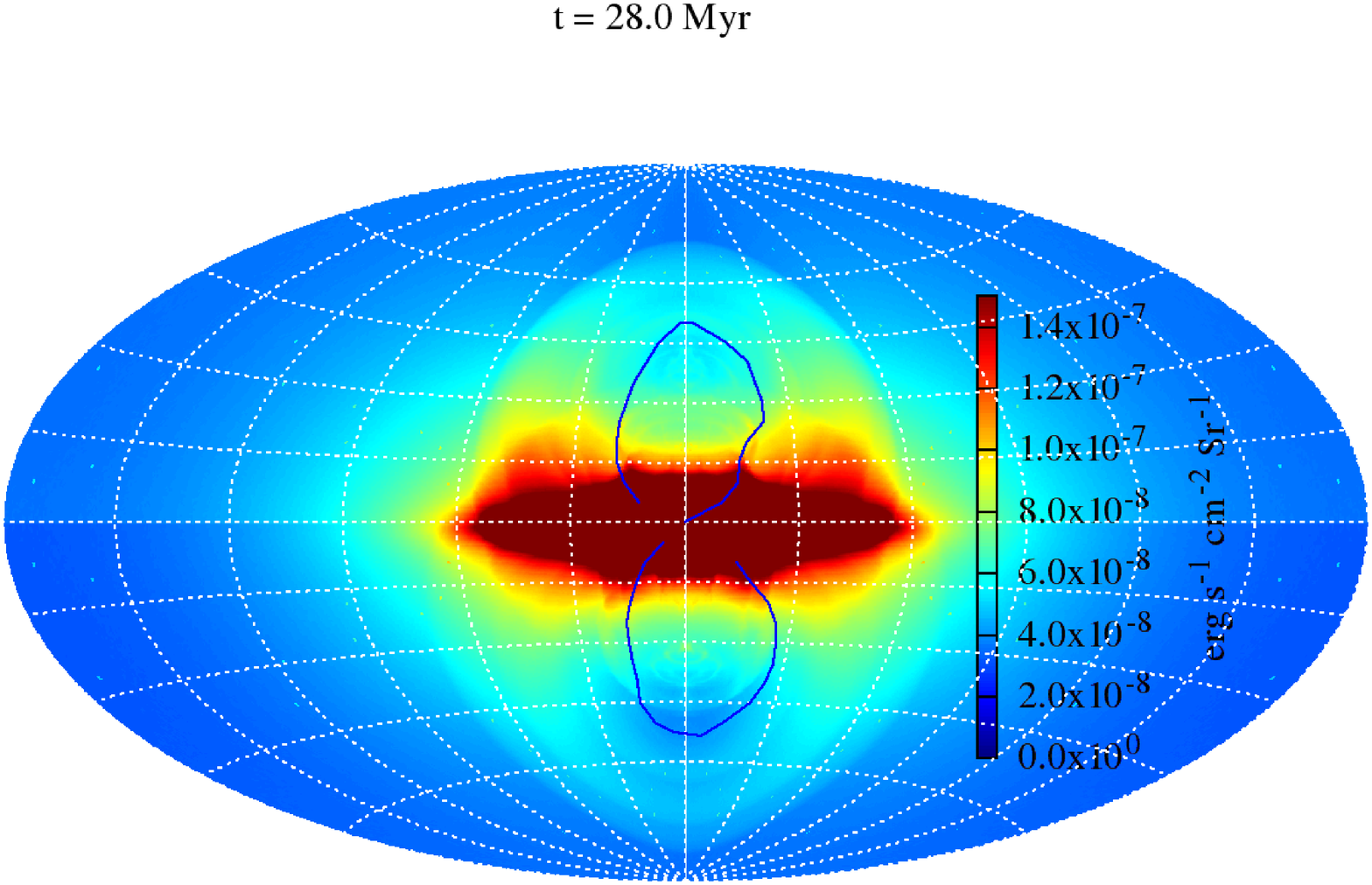}
	\caption{X-ray sky map at $0.5-2.0$ keV band for different CGM rotation values. From top to bottom, $f_h = 0, 1/3, 1/2$ and $2/3$. Although, an NPS like feature is not clearly visible in $f_h = 0, 2/3$ cases, the shell structure and their asymmetry can still be seen. The central bright patch is the X-ray emission due to interaction between the shock and the disc gas.}
\end{figure}
\vfill\null

\begin{figure}
	\centering
	\includegraphics[width=0.41\textwidth, clip=true, trim={6cm 2cm 6cm 3cm}]{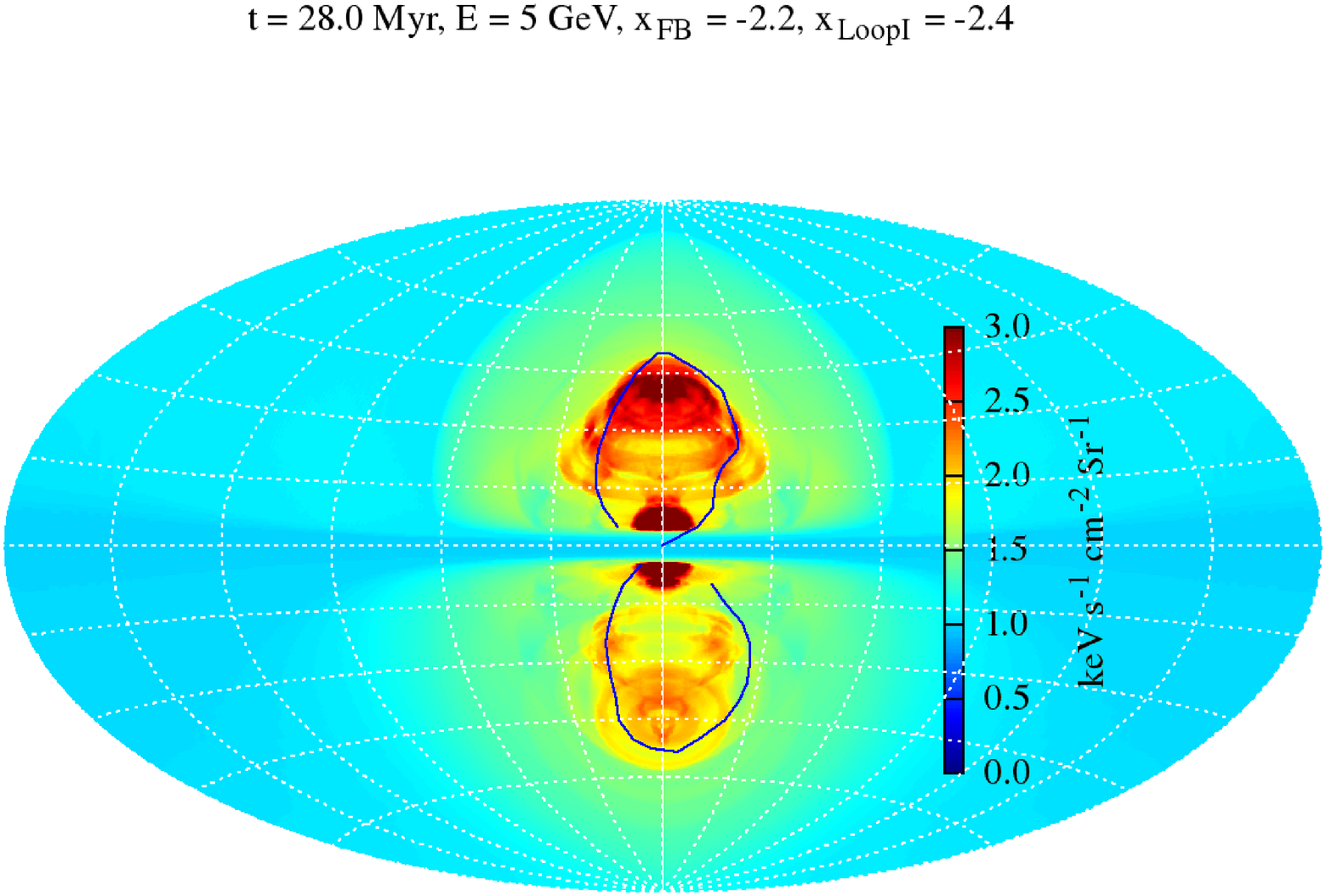}
	\includegraphics[width=0.41\textwidth, clip=true, trim={6cm 2cm 6cm 3cm}]{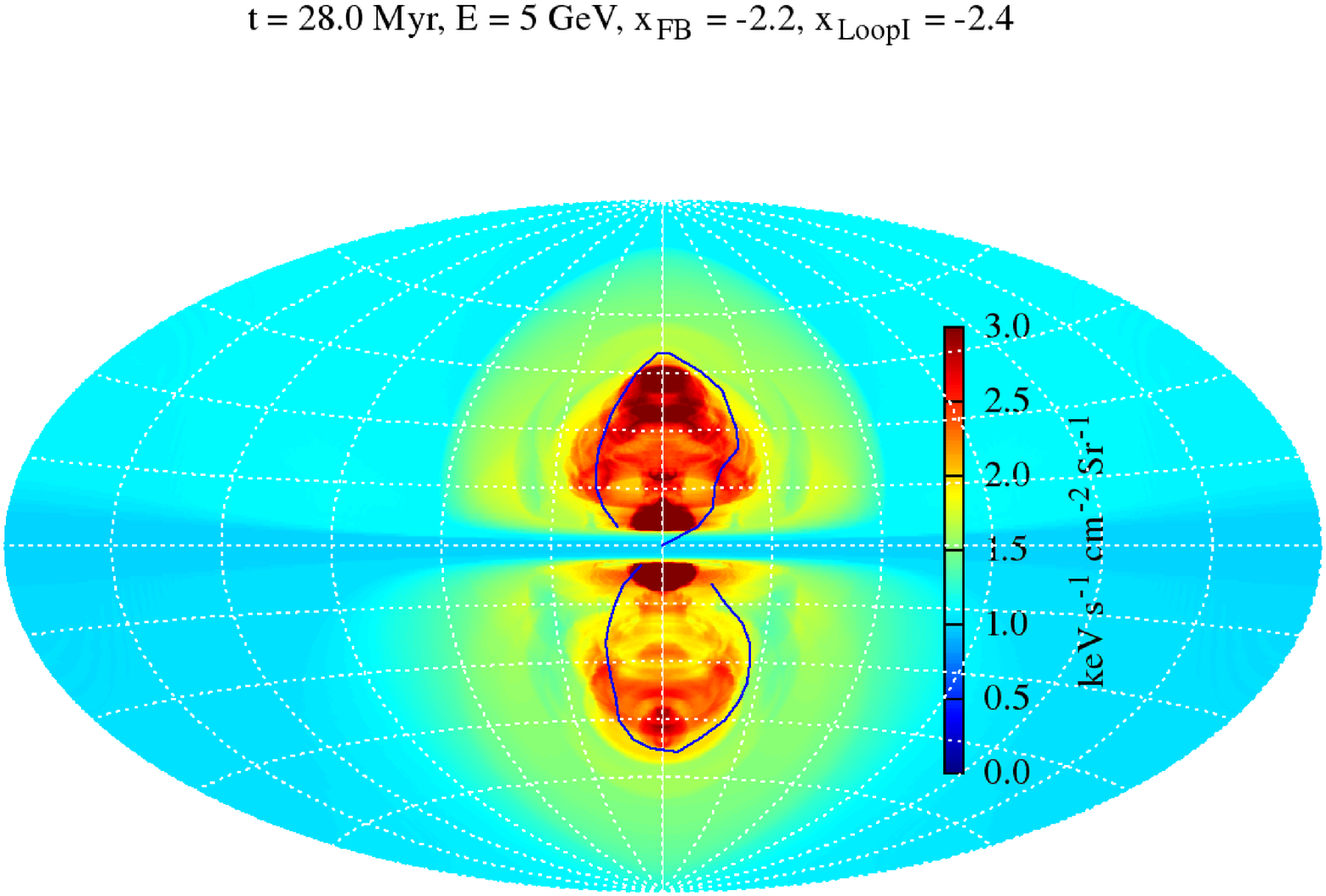}
	\includegraphics[width=0.41\textwidth, clip=true, trim={6cm 2cm 6cm 3cm}]{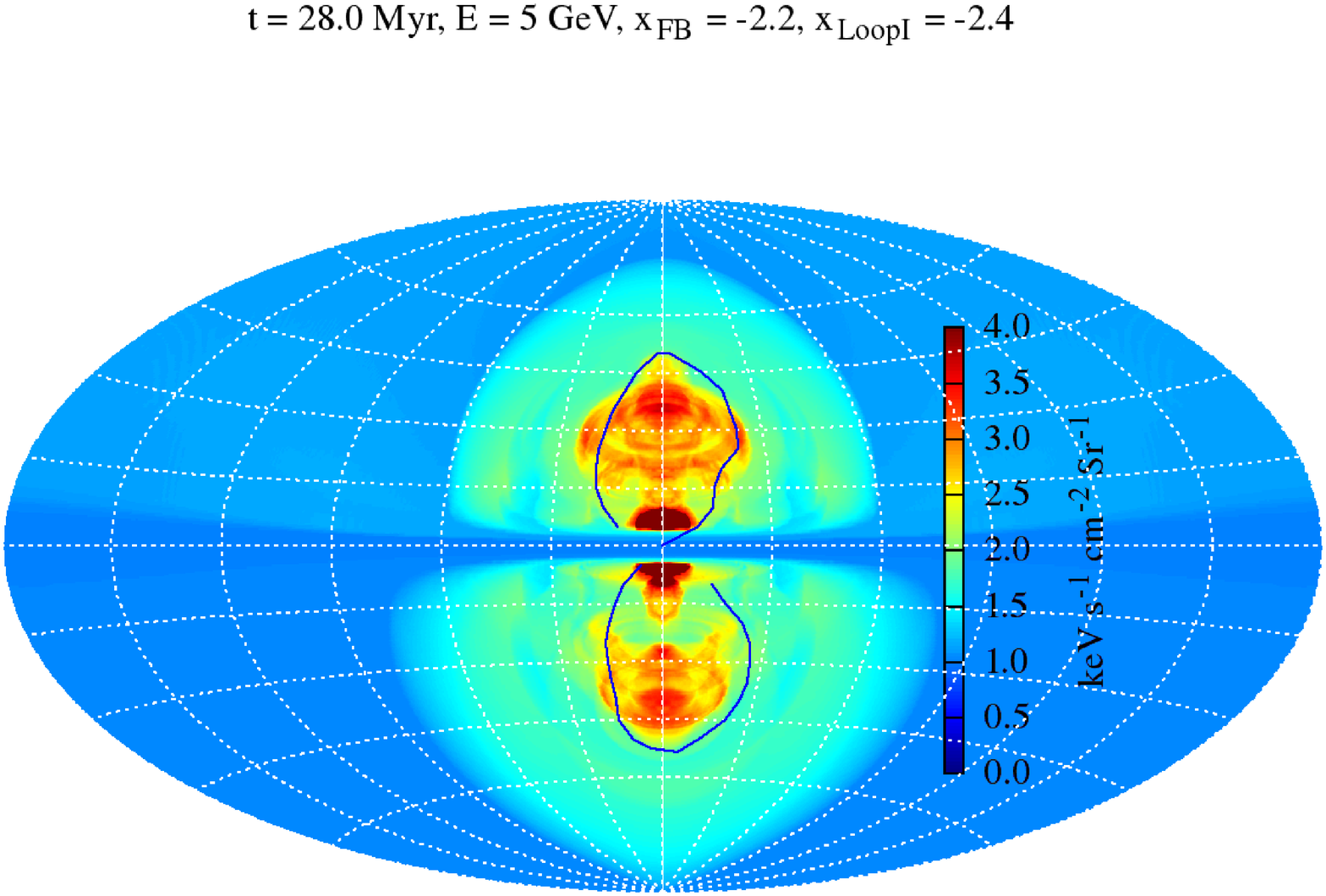}
	\includegraphics[width=0.41\textwidth, clip=true, trim={6cm 2cm 6cm 3cm}]{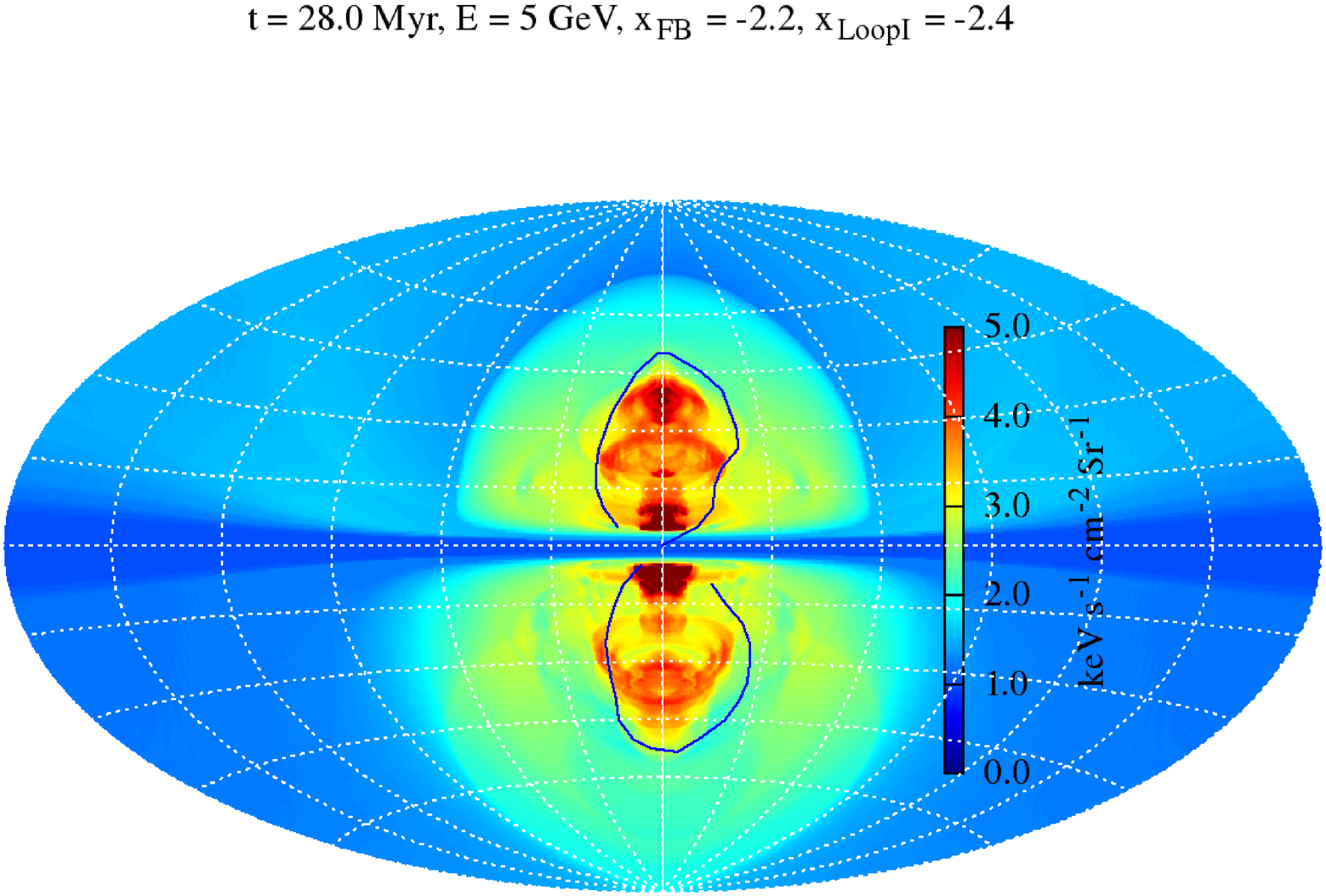}
	\caption{$\gamma$-ray sky map at $5$ GeV band for different CGM rotation values. From top to bottom, $f_h = 0, 1/3, 1/2$ and $2/3$.  Although the sizes of the Fermi Bubbles match with the observations, the surface brightness is not uniform throughout the surface. This is probably due to absence of proper evolution of cosmic ray energy density and diffusion.}
\end{figure}

\bsp	
\label{lastpage}
\end{document}